\documentclass[11pt]{article}
\usepackage{aaspp4,lscape}
\tighten
 
\def\gtsima{$\, \buildrel > \over \sim \,$}
\def\ltsima{$\, \buildrel < \over \sim \,$}
\def\simgt{\lower.5ex\hbox{\gtsima}}
\def\simlt{\lower.5ex\hbox{\ltsima}}
 
\begin{document}
 
\title{$K$-Band Calibration of the Red Clump Luminosity}
 
\bigskip
 
\author{David R.~Alves}
 
\affil{Space Telescope Science Institute, 
3700 San Martin Dr., Baltimore, MD 21218 \\
Email: {\tt alves@stsci.edu} }
 
\bigskip
\bigskip
 
\begin{abstract}
 
The average near-infrared ($K$-band) luminosity 
of 238 {\it Hipparcos} red clump giants 
is derived and then used
to measure the distance to the Galactic center.
These {\it Hipparcos} red clump giants
have been previously employed as $I$-band standard candles.
The advantage of the $K$-band 
is a decreased
sensitivity to reddening, and perhaps also a 
reduced systematic dependence on metallicity.
In order to
investigate the latter, and also to refer our calibration
to a known metallicity zero-point, 
we restrict our sample of red clump calibrators
to those with abundances derived from high-resolution
spectroscopic data.  
The mean metallicity of the sample
is [Fe/H]~=~$-0.18$~dex ($\sigma = 0.17$~dex).
The data
are consistent with no correlation between $M_{K}$ and [Fe/H],
and only weakly constrain the slope of this relation.
The luminosity function of the sample peaks
at $M_{K} = -1.61 \pm 0.03$~mag.
Next, we assemble published 
optical and near-infrared photometry for $\sim$ 20 red clump 
giants in a Baade's Window field with a mean metallicity
of [Fe/H]~=~$-0.17 \pm 0.09$~dex, which is
nearly identical to that of the {\it Hipparcos} red clump. 
Assuming that the average $(V-I)_{0}$ and $(V-K)_{0}$ colors
of these two red clumps are the same, the 
extinctions in the Baade's Window field 
are found to be $A_{V}$/$A_{I}$/$A_{K}$ = 1.56/0.87/0.15,
in agreement with previous estimates.
We derive the distance to the Galactic center: 
$(m-M)_{0} = 14.58 \pm 0.11$~mag, or 
$R = 8.24 \pm 0.42$~kpc.  The uncertainty in this 
distance measurement is dominated
by the small number of Baade's Window
red clump giants examined here.

\end{abstract}
 
\clearpage
\section{Introduction}

The horizontal branch ``red clump'' is
a landmark feature in color-magnitude diagrams of
intermediate-age ($\sim$1-10 Gyr) 
or very old ($\simgt$10 Gyr) but metal-rich stellar populations.
It has been suggested that red clump giants are standard candles, 
like their very old
and metal-poor cousins, the RR~Lyrae variable
stars (Cannon 1970).
Red clump giants have recently gained popularity as standard
candles because the {\it Hipparcos} color-magnitude diagram
of the solar neighborhood
displays a very prominent red clump (Perryman et al.~1997).
Paczy\'{n}ski and Stanek (1998) measured the distance to the
Galactic center using the {\it Hipparcos} $I$-band calibration of the
red clump luminosity.
Distance estimates to the
Large Magellanic Cloud (LMC) and M31 soon followed
(Udalski et al.~1998; Stanek et al.~1998; Stanek \& Garnavich 1998).
At first, the {\it Hipparcos} red clump yielded a
``short'' distance to the LMC, 
consistent with earlier measurements based on the red clump
(Butcher 1977; Seidel et al.~1987).  However, Romaniello et al.~(1999)
have since obtained a long LMC distance also using the red clump.
Extant distance measurements made with the LMC red clump 
do not agree.
Recent theoretical and observational studies of 
the effects of mean population age and metallicity on the
red clump luminosity have been made by Cole (1998),
Girardi et al.~(1998), Udalski (1998, 1998b, 1999), 
Alves and Sarajedini (1999), and others.
Earlier studies 
include Lattanzio (1986), Seidel et al.~(1987b), and
Olszewski et al.~(1987).
Variations in 
mean population age and metallicity are theoretically predicted to change
the red clump luminosity at the $\sim$20\% level.
Finally, issues of extinction that potentially affect 
recent distance measurements made with the
red clump have been discussed by
Romaniello et al.~(1999), Zaritsky~(1999), and Popowski~(2000).

In the spirit of McGonegal et al.~(1982), who first discussed the advantages
of the near-infrared wavelength regime for 
distance scale work with Cepheid variable stars, we assemble
$K$-band photometry data for 238 {\it Hipparcos} red clump
giants from the literature, 
derive a $K$-band luminosity calibration, and apply it
to published $K$-band photometry of Baade's Window in order to measure
the distance to the Galactic center.  The advantage of our
$K$-band calibration compared to the  
Paczy\'{n}ski and Stanek (1998) $I$-band
calibration is a decreased sensitivity to reddening
(e.g.~Cardelli et al.~1989),
and perhaps also a reduced systematic dependence on metallicity.  
In order to
investigate the latter, and also to refer our calibration 
to a known metallicity zero-point,
we restrict our sample of {\it Hipparcos} red clump calibrators to those
with abundances derived from high-resolution spectroscopic
data (McWilliams 1990).  Udalski (1999) made an important analysis
of the $M_{I}$--[Fe/H] correlation with these same data from McWilliams (1990).

\section{The Red Clump Data}

A.~Udalski kindly provided a machine-readable list of
{\it Hipparcos} catalog 
identifiers, ``Henry Draper Catalog''
names (Cannon \& Pickering 1918), 
and McWilliams (1990) metallicity estimates
for 284 red clump giants.
We downloaded
the {\it Hipparcos} main catalog from the NASA Astronomical
Data Center archive ({\tt http://adc.gsfc.nasa.gov/}) and assembled
RA \& Dec.~(H3-4),
parallax (H11), parallax error (H16), 
Johnson $V$ mag (H5), Johnson $(B-V)$ color (H37), 
and Cousins $(V-I)$ color (H40)
for each red clump giant in the Udalski list.
(Column entries in the {\it Hipparcos} main catalog
are given in parentheses.)

The coordinates of the 284 red clump calibrators 
were first cross-correlated with 
released 2MASS data using the NASA Infrared Science Archive
({\tt http://irsa.ipac.caltech.edu/}).  
Unfortunately, all 25 stars found carried saturation flags.
Therefore, we proceeded to download machine-readable versions of the
``Catalog of Infrared Observations, Edition 5''
(the ``CIO''; Gezari et al.~1999)
and the ``Infrared Source Cross Index'' (Schmitz et al.~1987) from
the NASA Astronomical Data Center archive.  The CIO
lists every astronomical photometric measurement 
in the wavelength range of 1 $\micron$ to 1 mm published since 1965.
Using the Henry Draper Catalog names of our red
clump calibrators, we retrieved the various different names for these
stars as they may appear in the CIO, and extracted all corresponding entries.
Listings of the 2.2 $\micron$ flux
with flags "M" or "C" (magnitude, or magnitude derived from color)
were then assembled.
Many of our 284 red clump calibrators were included in the
``Infrared Catalog'' (designated with ``IRC'' numbers), 
also known as the ``Caltech 
Two Micron Sky Survey'', or the ``TMSS'' (Neugebauer \& Leighton 1968, 1969).
We note that all other 2.2 $\micron$
($K$-band) measurements in the intervening 30 years 
showed remarkably good agreement 
with the TMSS data, typically to within 0.01 to 0.02 mag.  
As many as a dozen 2.2 $\micron$ measurements have been published for some
of our red clump calibrators, although
1-3 is typical.
Table~1 summarizes
(1) Henry Draper Catalog name, (2) other names for this star in
the CIO, (3) the average 2.2 $\micron$ mag for listings in the CIO under the
names given in column~2, (4) the {\it Hipparcos} identifier,
(5) the parallax ($\pi$) in units of mas,
(6) the parallax error ($\sigma_{\pi}$) in units of mas, (7) the $V$ mag,
(8) the ($B-V$) color, (9) the ($V-I$) color, and (10) the McWilliams (1990)
estimate of [Fe/H].  
We provide $K$-band magnitudes for 238 of the 284 red clump calibrators.

As a check of the $K$ photometry,
we identified 31 of our red clump calibrators in the 
Koorneef (1983) standard star list, which is primarily a compilation of 
Johnson and Glass standards.
The Johnson-Glass system has been thoroughly discussed
by Bessell and Brett (1988).  For the 31 red clump calibrators that are
also standard stars, the $K$ mags in Table~1 show no systematic
offset from the standard system, and a standard deviation of 0.014 mag.
Therefore, with regard to calculating the absolute magnitudes of the
red clump calibrators, the
errors associated with the published $K$-band photometry
are probably negligible compared to the uncertainties of
the {\it Hipparcos} parallaxes.

\section{The $K$-Band Calibration}

The red clump calibrators were preselected to have parallax
errors of less than 10\%
(Paczy\'{n}ski \& Stanek 1998).  The mean parallax error
of our sample is 5\%.
We calculate absolute magnitudes and errors
using the formulae,
\begin{equation}
M_{K}  =  K  -  10.0  +  2.1715 \cdot \ln(\pi)
\end{equation}
\begin{equation}
\sigma_{M_{K}}  = 2.1715 \cdot (\sigma_{\pi}/\pi)
\end{equation}
We assume no foreground reddening for our sample of red
clump calibrators (Mendez \& van Altena 1998).

We present the $M_{I}$, $(V-I)_{0}$ color-magnitude diagram
in Figure~1, which illustrates the color-magnitude cuts orginally
employed by Paczy\'{n}ski and Stanek (1998) to isolate the 
{\it Hipparcos} red clump.
Clearly, not all of these stars are red clump giants.
The $M_{K}$, $(V-K)_{0}$ color-magnitude diagram 
is shown in Figure~2.
The same stars are plotted in Figures~1 \& 2.
The primary concentration of red clump giants 
lies near $M_{K} \sim -1.6$ with colors
$(V-K)_{0} \sim 2.2$ to 2.5~mag.  The faint extension
of the red clump at $M_{K} \sim -1.2$ and
$(V-K)_{0} \sim 2.1$~mag has been discussed by
Girardi et al.~(1998).  The plume of brighter red clump stars
seen at $(V-K)_{0} \sim 2.1$~mag has been discussed
by Beaulieu and Sackett (1998).
Some stars with $(V-K)_{0} \simgt 2.5$~mag are probably
giant branch stars, not red clump giants.
Finally, the stars at $M_{K} \approx -2.85$ and
$(V-K)_{0} \sim 2.5$ to 3.0~mag 
are probably associated with
the asymptotic giant branch bump, and are not
red clump giants (e.g.~Gallart 1998,
Alves \& Sarajedini 1999).

In Figure~3, we plot the $M_{K}$ differential luminosity function of
the red clump calibrators.  
It has become customary to fit a linear
(or higher order) background plus a Gaussian model
to the red clump luminosity
function in order to derive the peak magnitude.  Inspection of Figure~3
suggests that a linear background is unnecessary, and we instead
allow for a constant background level in our fits.  Considering the 
likely contamination of the red clump calibrator sample (i.e,. by the
asymptotic giant branch bump at $M_{K} \approx -2.85$ mag), we 
fit the data in the restricted range of $-2.5 < M_{K} < -0.8$ mag.
This portion of the luminosity function is indicated  with a solid line
in Figure~3.  The excluded red clump data are shown with
a dotted line.  Our fit is a 3-parameter Gaussian (the center,
width, and height of the distribution) plus a constant background in
units of stars per histogram bin.  The best
fit yields $M_{K} = -1.61 \pm 0.03$~mag.  The width of the Gaussian component
in the model function is $\sigma_{K} = 0.22 \pm 0.03$ mag.  We report the
1$\sigma$ errors.  The best fit model function
is shown in Figure~3 with a dashed line.
Luminosity functions constructed with different 
magnitude cuts and histogram bin sizes 
yield consistent results for the center of the Gaussian; however,
the best-fit width of the Gaussian varies systematically at a level
comparable to the reported 1$\sigma$ error.
We note that the median value of the
red clump sample is $M_{K} = -1.62$ mag.  
If we apply a color cut of $2.2 < (V-K) < 2.5$,
the median value
is $M_{K} = -1.60$ mag.  These median values
reassure us that the ``wings'' of the peak in the $M_{K}$ luminosity function
(arising partly from contamination of the 
red clump sample by stars in other stellar evolutionary
phases) have not significantly affected our derivation of the
$K$-band red clump luminosity calibration.

In Figure~4,
we plot $M_{K}$ versus [Fe/H] for the {\it Hipparcos} 
red clump giants.
The mean metallicity of the red clump calibrators 
is [Fe/H]~=~$-0.18$~dex
($\sigma$~=~0.17 dex).
Using the errors from Eqn.~2, and adopting 
0.1~dex for the metallicity errors 
(McWilliams 1990),
the ordinary least squares
regression with [Fe/H] as the independent variable yields,
\begin{equation}
M_{K} \ = \ (0.57 \pm 0.36) \cdot {\rm [Fe/H]} \ - \ (1.64 \pm 0.07)
\end{equation}
A bootstrap analysis confirms the error estimate on the slope.
The inverse regression, the bisector regression, the orthogonal regression,
and the reduced major-axis regression (Isobe et al. 1990) yield wildly
discrepant results.  For comparison, if we apply a
color cut of $2.2 < (V-K) < 2.5$, the ordinary least squares
regression yields a slope of $-0.36 \pm 0.33$.
In conclusion, we do not find a significant correlation
between $M_{K}$ and [Fe/H].  These data only weakly constrain
the slope of a linear model dependence between $M_{K}$ and [Fe/H]
(i.e.~a range of slopes from $-0.36$ to 0.57 is found).
Indeed, a casual
inspection of Figure~4
suggests no dependence on metallicity.  We tentatively
recommend no metallicity correction to our $K$-band red clump
luminosity calibration and application only to red clump giants 
with metallicities in the restricted range of $-0.5 \simlt$
[Fe/H] $\simlt 0.0$.   Finally, we note
that these {\it Hipparcos} red clump data 
do not span a very wide range of metallicity,
which probably hinders our attempt to detect a correlation
between $M_{K}$ and [Fe/H].

\section{Distance to the Galactic Center}

Figure~5 shows a $K_{0}$, $(V-K)_{0}$ color-magnitude diagram
of Baade's Window.  The data are taken from Tiede et al.~(1995;
their field~4b), who give the dereddened magnitudes and
colors\footnote{The near-infrared magnitudes and colors
in Table 4 of Tiede et al.~(1995) are dereddened. 
However, the $V_{0}$ and
$(V-I)_{0}$ data are actually apparent magnitudes,
incorrectly labeled as dereddened magnitudes.}.
These authors adopt the following extinction coefficients for
this field: $A_{K}$ = 0.14, $A_{V}$ = 1.46, and $E(V-I)$ = 0.65 mag.
The photometric data presented in Figure~5 are dereddened 
with these coefficients.
Tiede et al.~(1995) find [Fe/H] = $-0.17 \pm 0.09$~dex 
for the mean metallicity of giants in this field,
which is based on their own analyses as well as a review of the
literature.
Thus, the mean metallicities of the Baade's Window and
{\it Hipparcos} red clumps 
examined here are nearly identical.  

The dereddened $K$-band brightness of the Baade's Window
red clump is calculated as follows.  In Figure~6,
we present a differential
luminosity function of the Baade's Window giant branch,
where we have excluded stars with $(V-K)_{0} < 3.0$,
$K_{0} < 10.5$, and $K_{0} > 15.0$ mag.  The adopted
bin size is 0.15 mag, which was determined after several
trials of model fitting, as described below.  Inspection of
Figure~6 suggests that a Gaussian plus a linear background
is an adequate model for the data.  
The best fit to the data yields
$<K_{0}>_{RC} = 12.98 \pm 0.11$ mag.
This is marked with an arrow in Figure~5.
The best-fit width of the
Gaussian component is $\sigma_{K} = 0.15$, although 
$\sigma_{K} = 0.22$ (as found for the {\it Hipparcos} red clump)
is consistent with the data.

In Figure~7, we present a dereddened color-color diagram.
The open circles are the
{\it Hipparcos} red clump giants with $-1.95 < M_{K} < -1.35$.
The filled circles are the Baade's Window red clump giants
with $12.7 < K_{0} < 13.2$.  We note again that these data 
are dereddened with
the extinction coefficients from Tiede et al.~(1995).
The solid line is the solar neighborhood giant sequence from
Bessell and Brett (1988; see also Fig.~14 of Tiede et al.~1995).
Finally, the reddening vector is shown in the lower right
($A_{K}$ = 0.01 mag).
All of the red clump giants lie along the fiducial
sequence, and therefore none would be considered anomalous.  
This is in contrast
to Baade's Window giants of later spectral types (i.e., those with
$(V-K)_{0} \simgt 3.0$~mag), which are
known to have anomalously red colors (Tiede et al.~1995). 
The mean colors of the Baade's Window and {\it Hipparcos} red clump giants
plotted in Figure~7 are $(V-K)_{0}$ = 2.47 and 2.34, and
$(V-I)_{0}$ = 1.09 and 1.00 mag, respectively.  
The mean colors of the Baade's Window red clump would shift
by  $\Delta(V-K)_{0} = -0.14$ and
$\Delta(V-I)_{0} = -0.07$ mag, if we adopt 
the ratios of $A_{V}$/$A_{K}$ and $A_{I}$/$A_{K}$ given
by Tiede et al.~(1995) and revise the Baade's Window extinction by
$\Delta A_{K} = +0.015$ mag.
In this case, the mean colors
of the Baade's Window and {\it Hipparcos} red clumps would be
very nearly identical.
However, given our number statistics, and
allowing for small uncertainties in the extinction ratios, we prefer a slightly
more conservative correction of $\Delta A_{K} = +0.01$ mag.
In summary, 
we adopt total extinctions of $A_{V}$/$A_{I}$/$A_{K}$ = 1.56/0.87/0.15,
consistent with other determinations (see discussion in Alcock et al.~1998),
in which case the mean colors of the Baade's Window and 
{\it Hipparcos} red clumps
agree to within a few percent.

Based on our analysis of the red clump colors, we revise the
dereddened $K$-band brightness of the Baade's Window
red clump to $<K_{0}>_{RC} = 12.97 \pm 0.11$ mag.
The true distance modulus to the Galactic center is then
$(m-M)_{0} = 14.58 \pm 0.11$~mag,
which corresponds to
$R = 8.24 \pm 0.42$~kpc.   
For comparison, 
Reid (1993) gives $R = 8.0 \pm 0.5$~kpc as the ``best''
mean of all measurements since 1974.  
We report only the statistical error, which is dominated by
the small number of the Baade's Window red clump giants
examined here.
The uncertainty in the extinction correction
contributes a negligible systematic error. 
We expect no systematic error from the effects of mean clump
metallicity, because the metallicities of the
calibrating red clump giants and those
in the Galactic bulge are the same.
The systematic error from the photometric calibration of
the Baade's Window $K$-band data
is $\sim$2\% (Tiede et al.~1995).
Paczy\'{n}ski and Stanek (1998) estimate that volume-sampling effects
in the Baade's Window data contribute a negligible 
$\sim$2\% systematic error.  Finally, systematic errors from
the effects of mean clump age, helium abundance, or selective
elemental enhancement are unknown.

We do not confirm the
``color discrepancy'' of $\sim$0.2~mag in $(V-I)_{0}$
between the Baade's Window and {\it Hipparcos} red clumps
found by Paczy\'{n}ski and Stanek (1998; see also Paczy\'{n}ski 1998).
Stanek et al.~(1999) and  Paczy\'{n}ski et al.~(1999) 
have since published new multi-color optical photometry 
of the red clump in Baade's Window.  As discussed by these authors,
the original color discrepancy arose partly from 
photometric calibration error. 
For example, Paczy\'{n}ski et al.~(1999)
now find that the Baade's Window red clump is $\sim$0.1~mag redder
in $(V-I)_{0}$
than the {\it Hipparcos} red clump. 
For the same field surveyed by
Tiede et al.~(1995),
Paczy\'{n}ski et al.~(1999) find a mean dereddened color 
of $(V-I)_{0} = 1.11$~mag.  They adopt $E(V-I) = 0.60$~mag, which
corresponds to a mean apparent color of $(V-I) = 1.71$~mag.  
For comparison, the Tiede et al.~(1995) data plotted in Figure~6 have
a mean apparent color of $(V-I) = 1.74$~mag.  Given these similar
apparent colors, our analysis of
the $(V-I)_{0}$, $(V-K)_{0}$ color-color diagram suggests that
the remaining $\sim$0.1~mag color discrepancy in $(V-I)_{0}$ found
by Paczy\'{n}ski et al.~(1999) 
may be reconciled with a small adjustment of the Baade's Window
extinction zero-point
(assuming standard extinction ratios).
We emphasize that uncertainties associated with the extinction correction
contribute a negligible systematic error to our $K$-band
distance measurement.

\section{Conclusion}

{\it Hipparcos} 
measured accurate parallaxes for several
hundred red clump
giants in the solar neighborhood. 
These stars may be useful standard candles. 
We have assembled $K$-band photometry data
from the literature for 238 of these {\it Hipparcos} red clump 
giants and calculated the mean absolute magnitude,
$M_{K} = -1.61 \pm 0.03$~mag.
The advantage of our
$K$-band calibration compared to the previously employed $I$-band
calibration is a decreased sensitivity to reddening,
and perhaps also a reduced systematic dependence on metallicity.
Using published $K$-band photometry of Baade's Window, we derive the
distance to the Galactic center, $R = 8.24 \pm 0.42$ kpc.  The error
in this distance measurement is dominated by the small number 
of Baade's Window red clump giants examined here.  Suggestions for
future work include refining the distance to the Galactic center by
using a larger sample of red clump giants with known metallicities,
reddenings, and $K$-band photometry.  Our $K$-band red clump 
calibration may also be used to determine the distance to the LMC, 
which currently serves as the zero-point for the extragalactic 
distance scale.


\section{Acknowledgements}
 
David Alves thanks Andrezej Udalski, Kris Stanek,
Howard Bond, Nino Panagia, and Nichole King 
for their useful comments on an early version of this paper.
The anonymous referee is thanked for several useful
remarks concerning the error analysis.
Support from the STScI Director's Discretionary
Research Fund is acknowledged.

\clearpage
\begin{figure}
\plotone{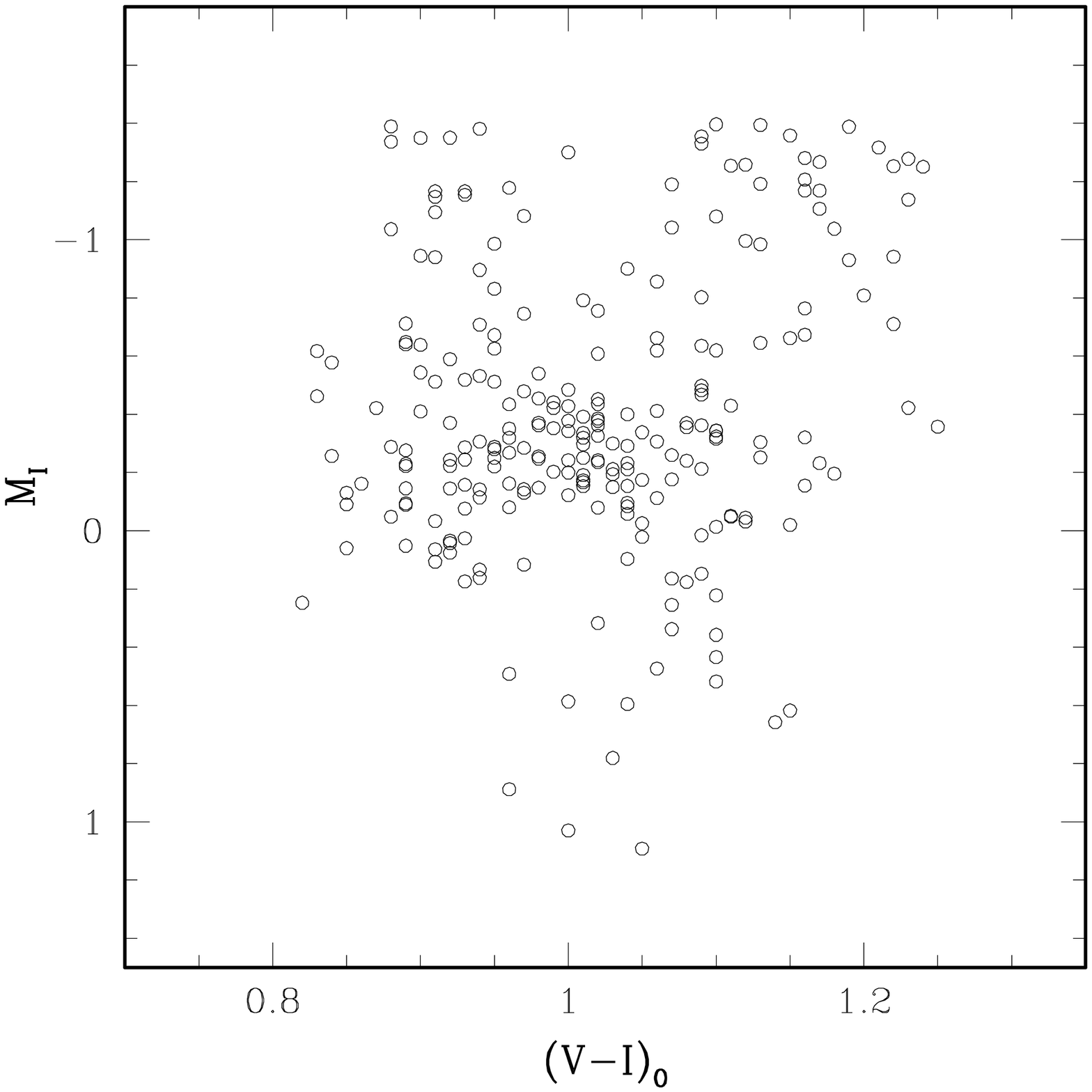}
\caption{$M_{I}$, $(V-I)_{0}$ color-magnitude diagram for 238 {\it Hipparcos}
red clump giants. The distribution of stars in this diagram reflects the
original color-magnitude cuts employed
by Paczy\'{n}ski and Stanek (1998).}
\end{figure}

\clearpage
\begin{figure}
\plotone{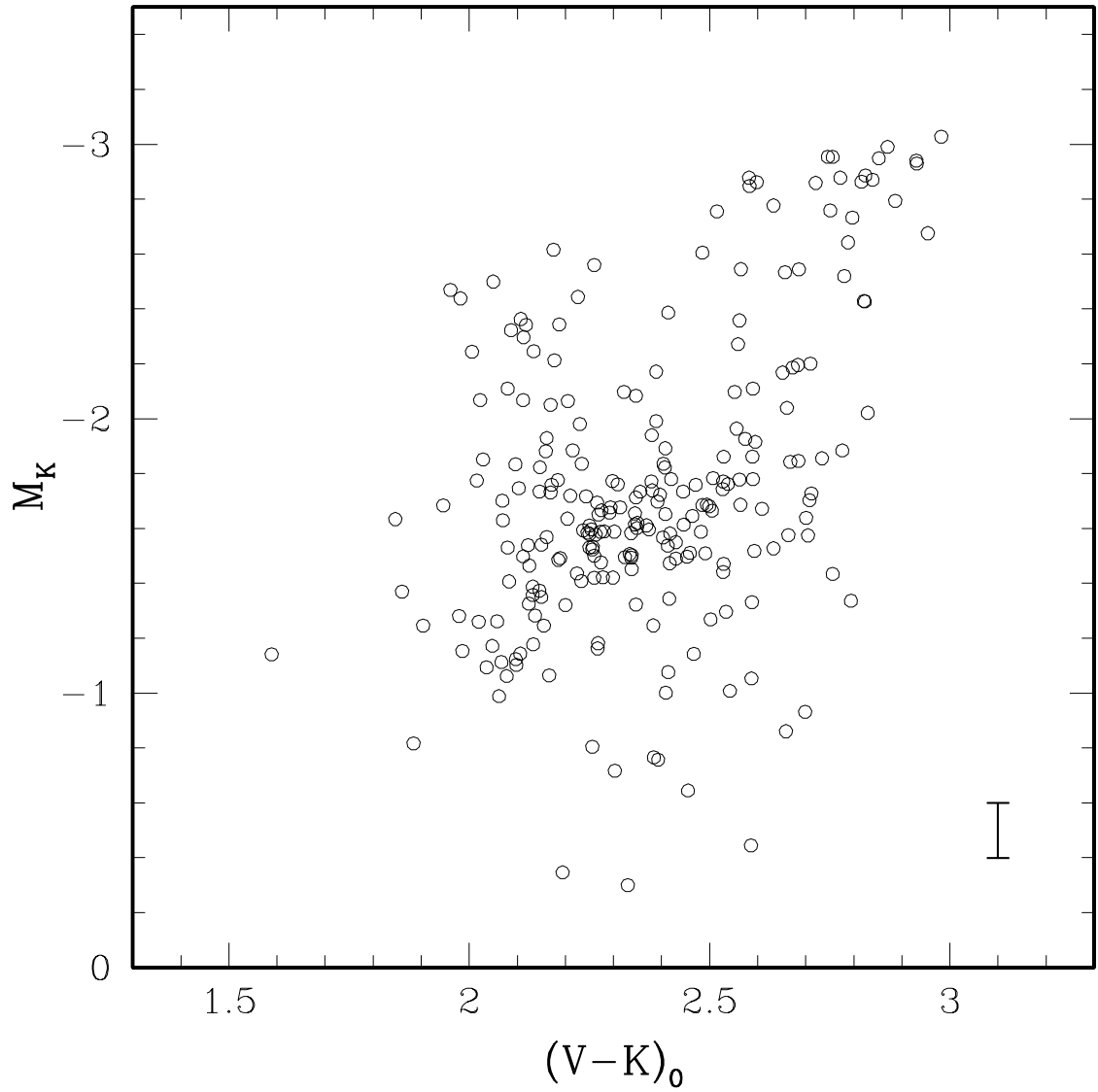}
\caption{$M_{K}$, $(V-K)_{0}$ color-magnitude diagram for 
the same 238 {\it Hipparcos} red clump giants plotted in Figure~1.
The average error bar 
is shown in the lower right.}
\end{figure}

\clearpage
\begin{figure}
\plotone{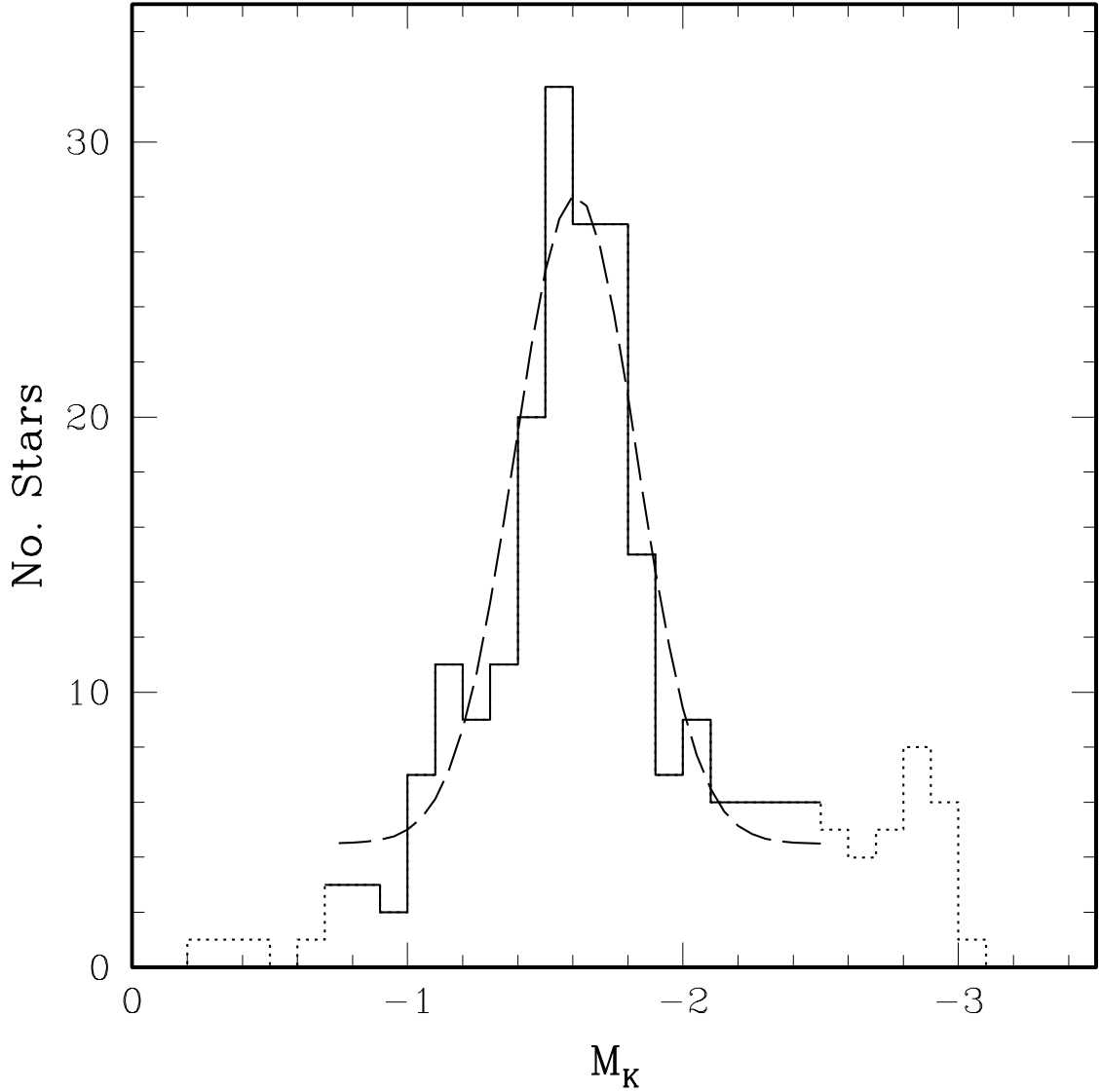}
\caption{The $M_{K}$
differential luminosity function for same 238 {\it Hipparcos}
red clump giants plotted in Figures 1 \& 2.  The histogram
bin size is 0.1 mag.
The region of the histogram used for fitting a Gaussian
plus a constant background model function is indicated with
a solid line.  The excluded regions are shown with a dotted
line.  The best fit model function is indicated with a dashed
line.  This fit yields our
$K$-band red clump luminosity calibration:
$M_{K}  = -1.61 \pm 0.03$~mag (see text for further details).}
\end{figure}

\clearpage
\begin{figure}
\plotone{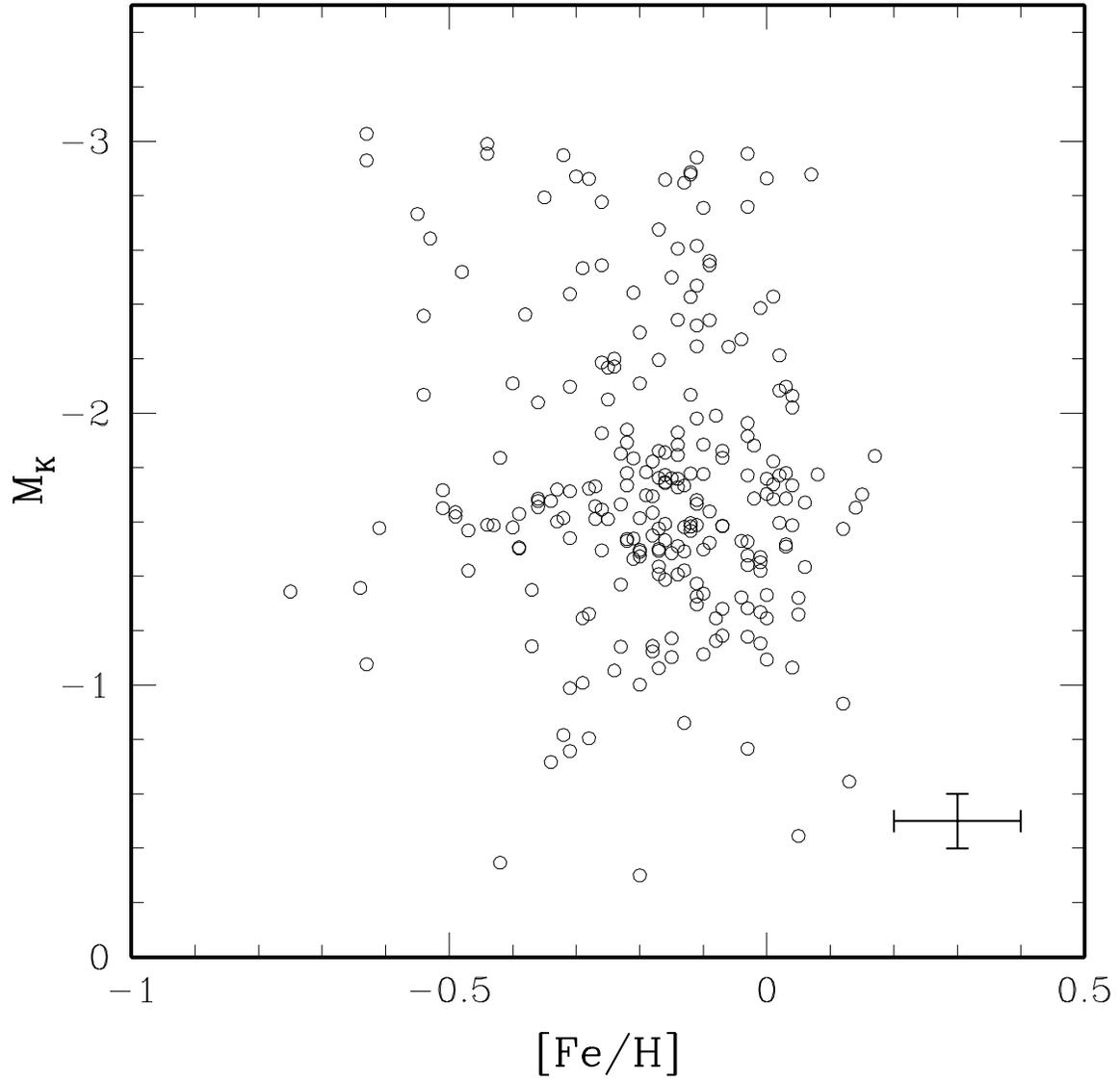}
\caption{$M_{K}$ versus [Fe/H] for the 238 {\it Hipparcos}
red clump giants (the same stars as shown in Figs.~1--3).  
Metallicity data are taken from McWilliams (1990).
The typical error bar is plotted in the lower right.}
\end{figure}

\clearpage

\begin{figure}
\plotone{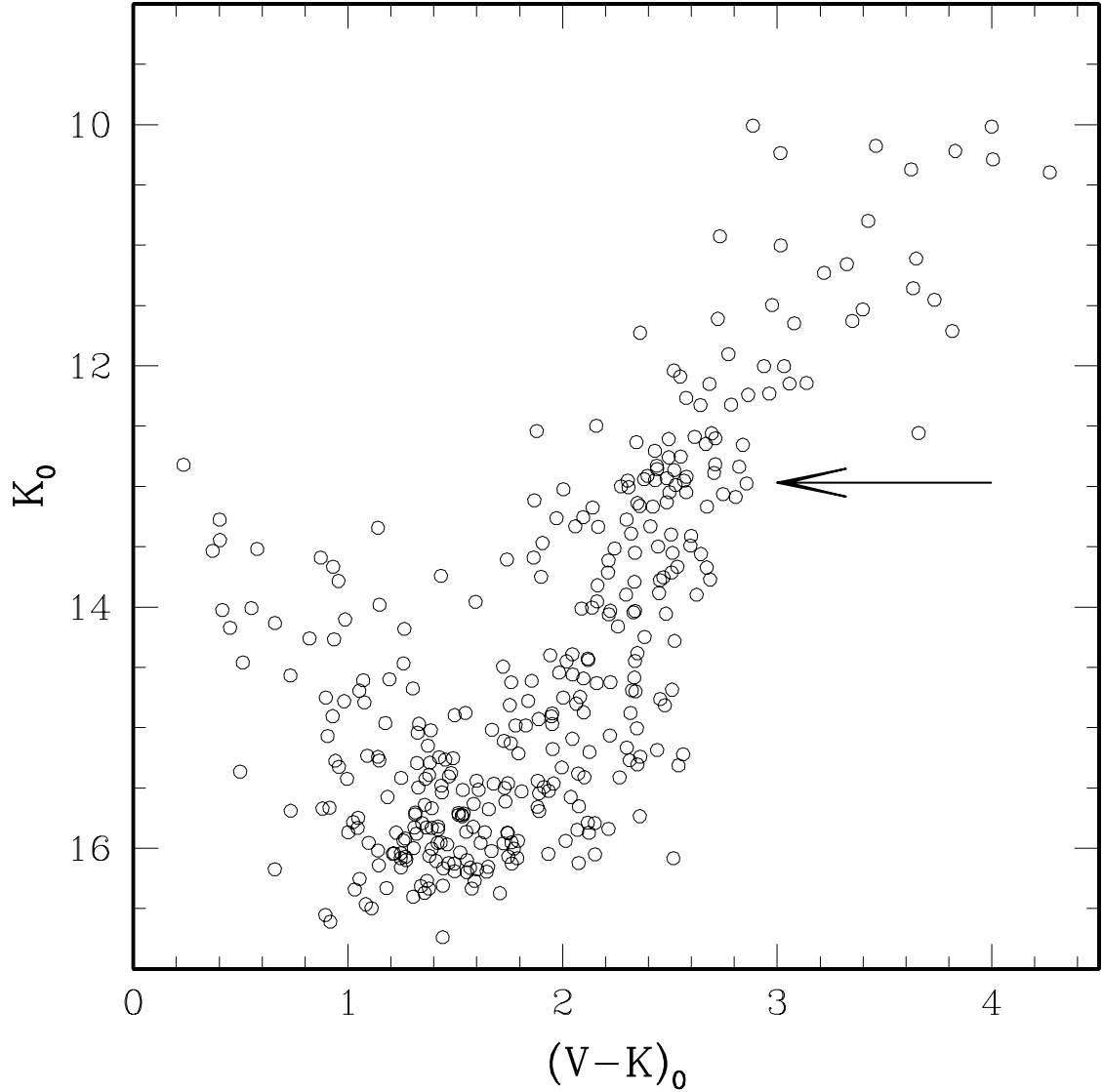}
\caption{$K_{0}$, $(V-K)_{0}$ color-magnitude diagram
for Baade's Window. Data are taken from 
Tiede et al.~(1995; their field~4b).  
The typical error bar is 0.05 mag at
$K_{0} \sim 13$ and 0.10 mag at $K_{0} \sim 16$.
The peak dereddened brightness of
the red clump, $K_{0} = 12.98 \pm 0.11$ mag, is
marked with an arrow (see also Fig.~6).  }
\end{figure}

\clearpage
\begin{figure}
\plotone{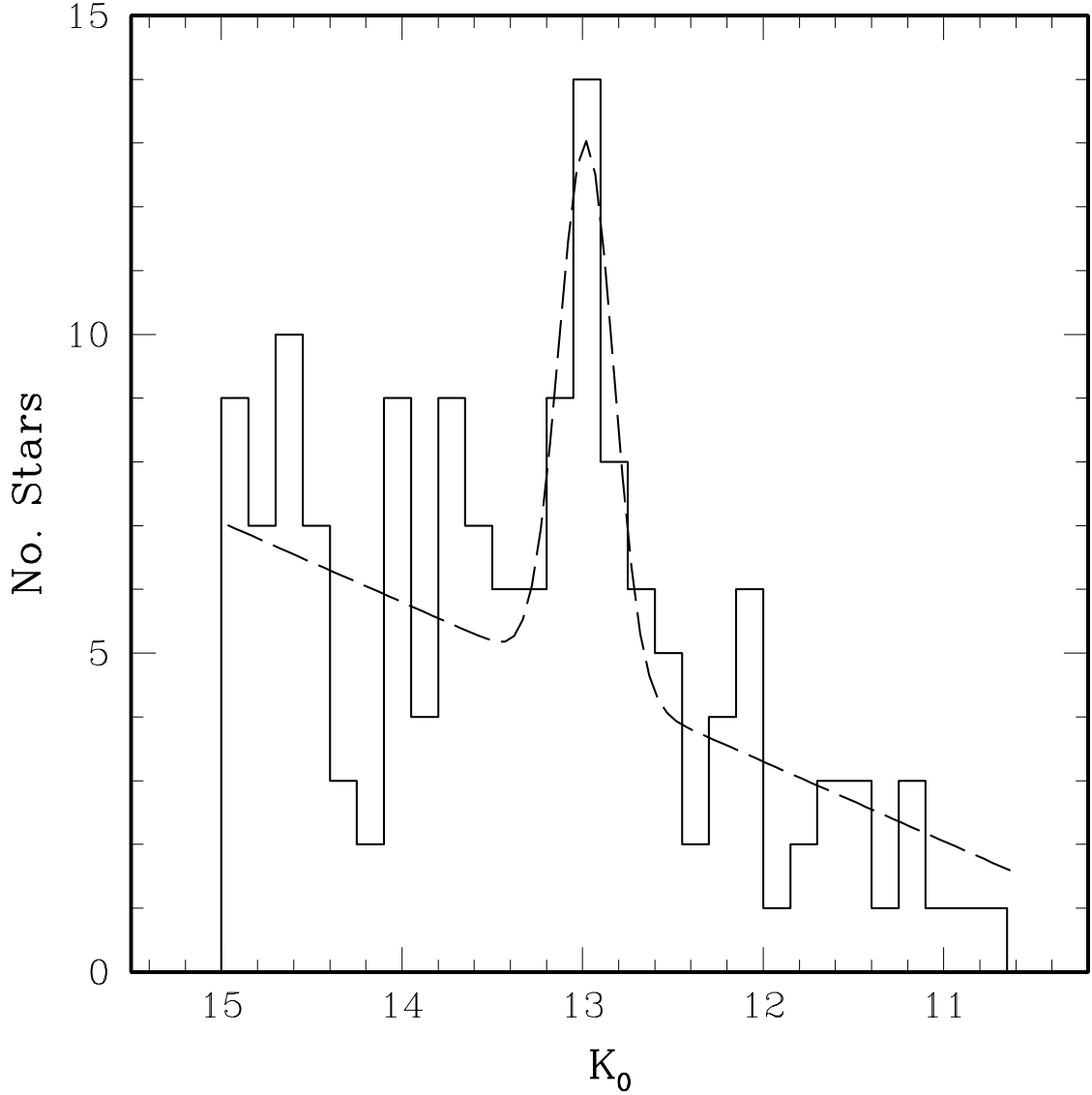}
\caption{The $K_{0}$ differential luminosity function for
the Baade's Window giant branch and red clump.  Some of the data
shown in Figure~5 have been excluded from this diagram, including
those stars with $(V-K)_{0} < 3.0$, $K_{0} < 10.5$, and 
$K_{0} > 15.0$ mag.  
The best-fit model function
consisting of a Gaussion and a linear background
is shown with a dashed line.  This fit yields
the peak dereddened brightness of the red clump,
$K_{0} = 12.98 \pm 0.11$ mag (see text for further details).}
\end{figure}

\clearpage
\begin{figure}
\plotone{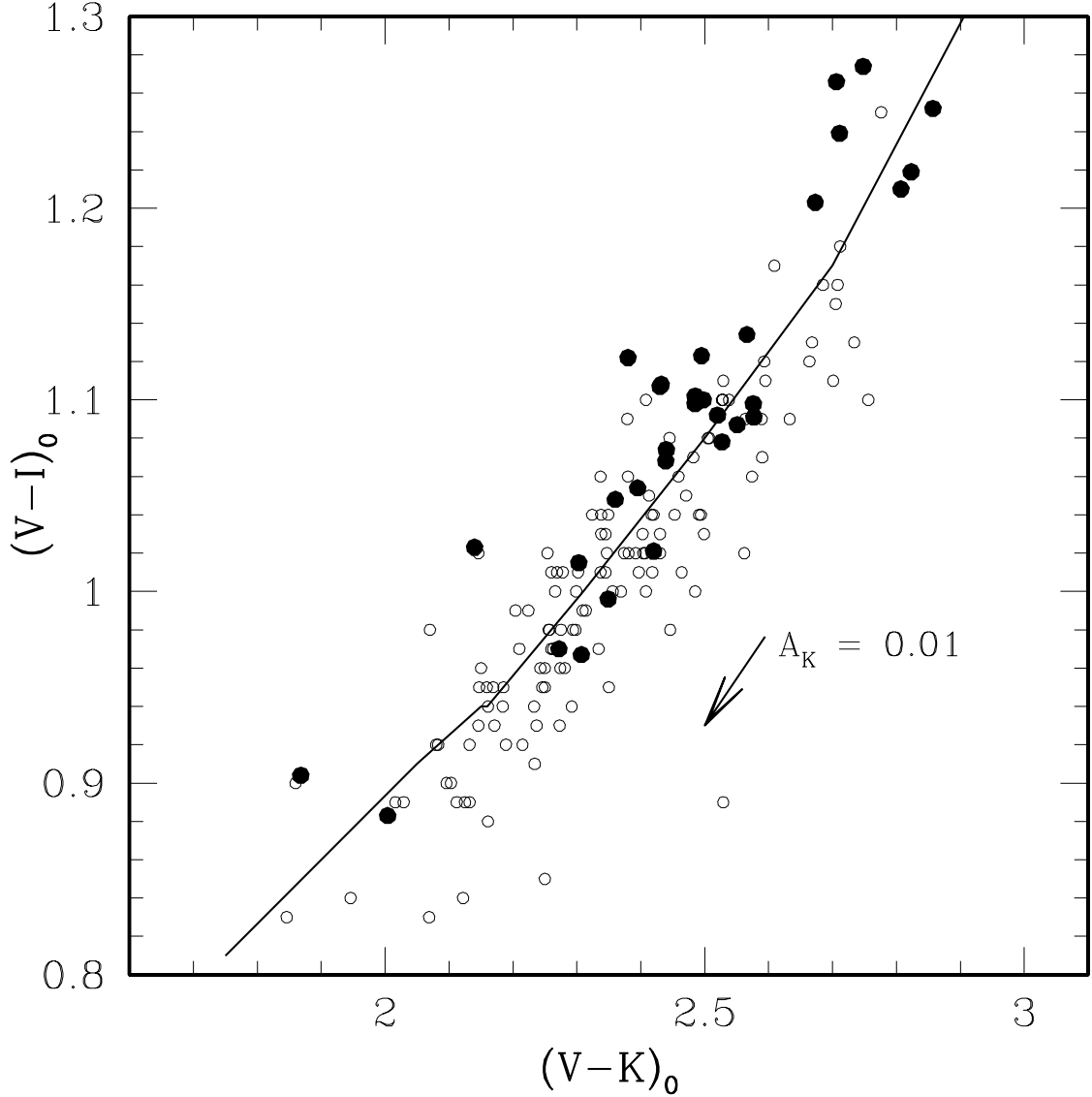}
\caption{The $(V-I)_{0}$, $(V-K)_{0}$ color-color diagram.
The open circles are the
{\it Hipparcos} red clump giants with $-1.95 < M_{K} < -1.35$.
The filled circles are the Baade's Window red clump giants
with $12.7 < K_{0} < 13.2$.
The line is the solar neighborhood giant sequence from
Bessell and Brett (1988).
The reddening vector, corresponding to $A_{K}$ = 0.01 mag, is
shown in the lower right.}
\end{figure}

\clearpage
\pagestyle{empty}
\textheight21cm
\textwidth12cm
\tighten
 
\begin{landscape}
\begin{deluxetable}{llrlrrrrrr}
\tablewidth20cm
\footnotesize
\tablecaption{Red Clump Giant Data }
\tablenum{1}
\tablehead{ 
\colhead{HD} & 
\colhead{CIO Names\tablenotemark{\ A} } &
\colhead{$K$}  &
\colhead{HIP.}  &
\colhead{$\pi$ {\footnotesize (mas)} } &
\colhead{$\sigma_{\pi}$ {\footnotesize (mas)} } &
\colhead{$V$}  &
\colhead{$B-V$}  &
\colhead{$V-I$}  &
\colhead{[Fe/H]}  
} 
\startdata 
    28 & IRC$-$10002, BS~3   & 2.22  &   443 & 25.38 & 1.05 &                4.613 &  1.029 &  1.04 & $-$0.31 \nl   
    87 &  &        &     476 &  8.75 & 0.80 &                                5.550 &  0.901 &  0.87 & $-$0.22 \nl   
   448 &     &     &     729 & 11.17 & 0.72 &                                5.570 &  1.043 &  0.99 & $-$0.01 \nl   
   645 &  &        &     873 & 15.35 & 0.74 &                                5.841 &  1.000 &  1.01 & $-$0.02 \nl   
  1737 &  &     &    1708 & 10.50 & 0.87 &                                   5.176 &  1.006 &  0.94 &  0.00 \nl     
  2910 &  &        &    2568 & 12.69 & 0.72 &                                5.377 &  1.069 &  1.00 & $-$0.08 \nl   
  3546 & IRC+30013, BS~163, $\epsilon$~And & 2.21 & 3031 & 19.34 & 0.76 &    4.342 &  0.871 &  0.92 & $-$0.64 \nl   
  3627 & IRC+30014, BS~165, $\delta$~And & 0.44  &  3092 & 32.19 & 0.68 &    3.269 &  1.268 &  1.23 &  0.04 \nl     
  3817 &    &   &    3231 &  9.47 & 0.81 &                                   5.302 &  0.891 &  0.91 & $-$0.26 \nl   
  4128 & IRC$-$20010, BS~188  & $-$0.22    &    3419 & 34.04 & 0.82 &        2.040 &  1.019 &  1.00 & $-$0.09 \nl   
  4188 & IRC$-$10012 & 2.51   &    3455 & 15.54 & 0.82 &                     4.767 &  0.998 &  0.98 & $-$0.16 \nl   
  5395 & IRC+60030 & 2.35     &    4422 & 15.84 & 0.58 &                     4.619 &  0.957 &  1.01 & $-$0.51 \nl   
  5516 & IRC+20015 & 2.29     &    4463 & 13.44 & 0.75 &                     4.402 &  0.940 &  0.94 & $-$0.54 \nl   
  5722 &    &     &    4587 & 10.35 & 0.96 &                                 5.615 &  0.949 &  0.96 & $-$0.37 \nl   
  6186 & IRC+10009, BS~294    & 2.20     &    4906 & 17.14 & 0.81 &          4.270 &  0.952 &  0.98 & $-$0.39 \nl   
  6805 & IRC$-$10018, BS~334  & 0.87   &    5364 & 27.73 & 0.71 &            3.465 &  1.161 &  1.11 & $-$0.03 \nl   
  7106 & IRC+30022    & 2.16     &    5586 & 20.11 & 0.87 &                  4.507 &  1.092 &  1.05 & $-$0.04 \nl   
  8207 & IRC+50033     & 2.38   &    6411 & 16.68 & 0.70 &                   4.871 &  1.077 &  1.04 &  0.03 \nl     
  8512 & IRC$-$10021, BS~402  & 1.19     &    6537 & 28.48 & 0.77 &          3.603 &  1.065 &  1.05 & $-$0.22 \nl   
  8491 & IRC+70027 & 2.44    &    6692 & 16.89 & 0.56 &                      4.718 &  1.047 &  1.01 & $-$0.13 \nl   
  8763 &  &                  &    6732 & 10.63 & 0.77 &                      5.503 &  1.106 &  1.04 & $-$0.12 \nl   
  9057 & IRC+50038     & 2.97     &    6999 & 11.26 & 0.77 &                 5.268 &  0.999 &  0.98 & $-$0.16 \nl   
  9408 & IRC+60057 & 2.33    &    7294 & 15.96 & 0.68 &                      4.675 &  0.991 &  1.01 & $-$0.36 \nl   
  9927 & IRC+50041, BS~464     & 0.77    &    7607 & 18.76 & 0.74 &          3.586 &  1.275 &  1.23 &  0.00 \nl     
 10072 & IRC+40025  & 2.89       &    7719 & 13.46 & 0.76 &                  5.015 &  0.883 &  0.89 & $-$0.21 \nl   
 10537 & IRC$-$30018  & 2.87     &    7955 & 15.02 & 0.72 &                  5.253 &  1.044 &  1.04 & $-$0.29 \nl   
 10761 & IRC+10021, BS~510     & 2.17    &    8198 & 12.63 & 0.86 &          4.257 &  0.942 &  0.93 & $-$0.11 \nl   
 11559 & IRC~~00027  & 2.46       &    8833 & 17.11 & 0.77 &                 4.606 &  0.928 &  0.93 & $-$0.11 \nl   
 12438 &     &               &    9440 & 11.71 & 0.73 &                      5.343 &  0.883 &  0.94 & $-$0.59 \nl   
 12339 &    &                &    9763 &  7.65 & 0.52 &                      5.218 &  0.954 &  0.94 & $-$0.10 \nl   
 12929 & IRC+20038, BS~617, $\alpha$~Ari & $-$0.64 & 9884 & 49.48 & 0.99 &   2.012 &  1.151 &  1.13 & $-$0.25 \nl   
 14129 &    &         &   10642 &  9.58 & 0.93 &                             5.507 &  0.962 &  0.93 & $-$0.10 \nl   
 15779 &     &               &   11791 & 12.27 & 1.13 &                      5.364 &  1.004 &  0.96 & $-$0.17 \nl   
 17361 & IRC+30050, BS~824     & 2.03   &   13061 & 18.06 & 0.84 &           4.524 &  1.112 &  1.04 & $-$0.02 \nl   
 17652 & IRC$-$30026, BS~841   & 2.15   &   13147 & 19.31 & 0.67 &           4.449 &  0.981 &  1.00 & $-$0.47 \nl   
 17824 & IRC$-$20037, BS~850     & 2.68   &   13288 & 17.85 & 0.69 &         4.758 &  0.906 &  0.91 & $-$0.17 \nl   
 18322 & IRC$-$10043, BS~874     & 1.39     &   13701 & 24.49 & 0.72 &       3.895 &  1.088 &  1.08 & $-$0.23 \nl   
 18970 & IRC+60109     & 2.55  &   14382 & 15.95 & 0.76 &                    4.774 &  1.018 &  0.99 & $-$0.17 \nl   
 19476 & IRC+40057, BS~941, $\kappa$~Per & 1.62  & 14668 & 29.05 & 0.66 &    3.786 &  0.980 &  0.94 & 0.04 \nl      
 19656 & IRC+40058 & 2.10   &   14817 & 10.69 & 0.80 &                       4.615 &  1.115 &  1.09 & $-$0.10 \nl   
 19787 & IRC+20054, BS~951  & 2.08     &   14838 & 19.44 & 1.23 &            4.354 &  1.033 &  0.96 & $-$0.03 \nl   
 20610 & IRC$-$20040     & 2.63     &   15382 & 12.79 & 0.75 &               4.864 &  0.904 &  0.91 & $-$0.07 \nl   
 20559 & IRC~~00044     & 2.83       &   15383 & 14.68 & 0.96 &              5.624 &  1.050 &  1.10 & $-$0.10 \nl   
 21017 & IRC+20059     & 2.91   &   15861 & 14.18 & 0.98 &                   5.498 &  1.190 &  1.08 &  0.00 \nl     
 21120 & IRC+10044, BS~1030  & 1.56     &   15900 & 15.42 & 1.28 &           3.610 &  0.887 &  0.90 & $-$0.15 \nl   
 22409 &    &                &   16780 &  8.60 & 0.77 &                      5.563 &  0.915 &  0.93 & $-$0.30 \nl   
 23940 &    &      &   17738 & 11.55 & 0.65 &                                5.523 &  0.973 &  0.97 & $-$0.43 \nl   
 25604 & IRC+20070, BS~1256   & 1.98     &   19038 & 18.04 & 0.84 &          4.361 &  1.064 &  1.02 &  0.01 \nl     
 26162 &    &                &   19388 & 11.21 & 0.87 &                      5.508 &  1.077 &  1.05 & $-$0.02 \nl   
 26409 &    &                &   19483 &  8.65 & 0.82 &                      5.445 &  0.941 &  0.94 & $-$0.10 \nl   
 27371 & IRC+20074, BS~1346, $\gamma$~Tau &  1.49 & 20205 & 21.17 & 1.17 &   3.649 &  0.981 &  0.95 & $-$0.02 \nl   
 27022 & BS~1327    & 3.41    &   20266 &  9.80 & 0.67 &                     5.256 &  0.820 &  0.83 & $-$0.18 \nl   
 27697 & IRC+20076, BS~1373  $\delta$ Tau & 1.60 & 20455 & 21.29 & 0.93 &    3.771 &  0.983 &  0.93 &  0.00 \nl     
 28292 & IRC+20079, BS~1407    & 2.30    &   20877 & 16.78 & 1.02 &          4.964 &  1.137 &  1.12 & $-$0.17 \nl   
 28307 & IRC+20081, BS~1411, $\theta^1$~Tau & 1.69 & 20885 & 20.66 & 0.85 &  3.836 &  0.952 &  1.02 &  0.04 \nl     
 28305 & IRC+20080, BS~1409, $\epsilon$~Tau & 1.32 & 20889 & 21.04 & 0.82 &  3.525 &  1.014 &  1.04 & 0.04 \nl      
 29085 & IRC$-$30036, BS~1453  & 2.19    &   21248 & 26.22 & 0.71 &          4.493 &  0.972 &  1.00 & $-$0.34 \nl   
 29291 & IRC$-$30037, BS~1464  & 1.69    &   21393 & 15.62 & 0.63 &          3.808 &  0.957 &  0.93 & $-$0.09 \nl   
 29503 & IRC$-$10073, BS~1481  & 1.33    &   21594 & 29.84 & 0.62 &          3.864 &  1.082 &  1.09 & $-$0.11 \nl   
 29613 & IRC$-$10074  & 2.87   &   21685 & 16.42 & 0.69 &                    5.457 &  1.054 &  1.06 & $-$0.24 \nl   
 30814 & IRC$-$20063  & 2.78   &   22479 & 13.40 & 0.68 &                    5.026 &  0.992 &  0.95 & $-$0.07 \nl   
 31421 & IRC+10075, BS~1580  & 1.39      &   22957 & 19.26 & 1.05 &          4.063 &  1.158 &  1.16 & $-$0.26 \nl   
 32436 & IRC$-$30041   & 2.76    &   23430 & 13.45 & 0.59 &                  5.014 &  1.056 &  1.02 &  0.02 \nl     
 34559 & IRC+20105  & 2.89     &   24822 & 15.83 & 0.86 &                    4.957 &  0.937 &  0.92 & $-$0.10 \nl   
 35369 & IRC$-$10089, BS~1784, 29~Ori & 1.92  &   25247 & 18.71 & 0.74 &     4.130 &  0.943 &  0.97 & $-$0.33 \nl   
 35410 & IRC~~00072  & 2.81     &   25282 & 18.93 & 0.82 &                   5.066 &  0.961 &  0.96 & $-$0.28 \nl   
 37160 & IRC+10091, BS~1907, 40~Ori  & 1.68    &   26366 & 28.10 & 0.91 &    4.094 &  0.951 &  1.02 & $-$0.63 \nl   
 37984 & IRC~~00083, BS~1963   & 2.10   &   26885 & 10.80 & 0.81 &           4.897 &  1.144 &  1.17 & $-$0.55 \nl   
 38656 & IRC+40142, BS~1995     & 2.34   &   27483 & 15.34 & 0.80 &          4.509 &  0.949 &  0.95 & $-$0.27 \nl   
 39364 & IRC$-$20081, BS~2035     & 1.34  &   27654 & 29.05 & 0.62 &         3.756 &  0.984 &  1.05 & $-$0.75 \nl   
 39003 & IRC+40144, BS~2012, $\nu$~Aur & 1.49  &   27673 & 15.17 & 0.88 &    3.975 &  1.132 &  1.07 & $-$0.14 \nl   
 40035 & IRC+50155, BS~2077  & 1.41     &   28358 & 23.22 & 0.91 &           3.719 &  1.010 &  0.99 & $-$0.15 \nl   
 41116 & IRC+20131, BS~2134  & 2.17     &   28734 & 21.64 & 1.06 &           4.156 &  0.835 &  0.88 & $-$0.01 \nl   
 41597 & IRC+60162     & 2.79  &   29246 &  9.34 & 0.69 &                    5.352 &  1.096 &  1.06 & $-$0.54 \nl   
 43039 & IRC+30146, BS~2219  & 1.97     &   29696 & 19.31 & 0.83 &           4.319 &  1.021 &  1.04 & $-$0.33 \nl   
 44762 & BS~2296  & 1.87   &   30277 & 13.75 & 0.60 &                        3.852 &  0.858 &  0.88 & $-$0.31 \nl   
 48433 & IRC+10134, BS~2478  & 1.86     &   32249 & 11.82 & 0.83 &           4.493 &  1.167 &  1.11 & $-$0.26 \nl   
 50522 & IRC+60178, BS~2560, 15~Lyn & 2.33    &   33449 & 19.14 & 0.76 &     4.350 &  0.850 &  0.85 &  0.05 \nl     
 58207 & IRC+30183, BS~2821   & 1.44     &   36046 & 25.90 & 0.91 &          3.777 &  1.024 &  1.01 & $-$0.17 \nl   
 61935 & IRC$-$10174, BS~2970  & 1.64    &   37447 & 22.61 & 0.80 &          3.942 &  1.022 &  1.01 & $-$0.11 \nl   
 62345 & IRC+20188, BS~2985  & 1.47     &   37740 & 22.73 & 0.83 &           3.573 &  0.932 &  0.90 & $-$0.16 \nl   
 62509 & IRC+30194, BS~2990, $\beta$~Gem & $-$1.11 & 37826 & 96.74 & 0.87 &  1.158 & 0.991 & 0.97 & $-$0.07 \nl     
 65695 & IRC~~00165   & 2.22     &   39079 & 13.06 & 0.96 &                  4.930 &  1.205 &  1.22 & $-$0.24 \nl   
 66216 & IRC+30198  & 2.39  &   39424 & 12.66 & 0.78 &                       4.942 &  1.130 &  1.09 &  0.03 \nl     
 68290 & IRC$-$10188, BS~3211  & 2.59   &   40084 & 17.64 & 0.69 &           4.723 &  0.939 &  0.93 & $-$0.03 \nl   
 71115 &    &      &   41325 &  9.08 & 0.87 &                                5.130 &  0.934 &  0.93 & $-$0.34 \nl   
 73898 & IRC$-$30131    & 2.66     &   42483 & 13.83 & 0.52 &                4.864 &  0.900 &  0.99 & $-$0.49 \nl   
 73108 & IRC+60188, BS~3403  & 1.90     &   42527 & 12.92 & 0.71 &           4.586 &  1.179 &  1.18 & $-$0.26 \nl   
 74137 & IRC$-$20174    & 2.53     &   42662 & 15.98 & 0.80 &                4.868 &  1.063 &  1.04 & $-$0.01 \nl   
 74442 & IRC+20205, BS~3461  & 1.52     &   42911 & 23.97 & 0.83 &           3.938 &  1.083 &  1.01 & $-$0.13 \nl   
 75691 & IRC$-$30136, BS~3518  & 1.09   &   43409 & 15.63 & 0.58 &           4.020 &  1.272 &  1.24 & $-$0.11 \nl   
 76294 & IRC+10196, BS~3547, $\zeta$~Hya & 0.88 & 43813 & 21.64 & 0.99 &     3.106 &  0.978 &  0.96 & $-$0.21 \nl   
 80499 & IRC$-$10214  & 2.66    &   45751 & 10.20 & 0.83 &                   4.773 &  0.927 &  0.91 & $-$0.20 \nl   
 80586 & IRC$-$10215    & 2.78    &   45811 & 13.39 & 0.82 &                 3.936 &  4.804 &  0.92 & $-$0.07 \nl  
 81169 & IRC$-$30151, BS~3733  & 2.61    &   46026 & 17.92 & 0.73 &          4.707 &  0.892 &  0.91 & $-$0.18 \nl   
 81146 & IRC+30211, BS~3731  & 1.65   &   46146 & 15.28 & 0.83 &             4.471 &  1.222 &  1.20 &  0.01 \nl     
 81799 & IRC$-$20190  & 2.19     &   46371 & 18.53 & 1.10 &                  4.719 &  1.154 &  1.11 & $-$0.01 \nl   
 82395 & IRC+10206  & 2.46     &   46771 & 13.67 & 0.87 &                    4.989 &  1.046 &  0.89 & $-$0.17 \nl   
 82635 & IRC+40208  & 2.49     &   46952 & 18.52 & 0.88 &                    4.538 &  0.914 &  0.91 & $-$0.15 \nl   
 82741 & IRC+40209   & 2.54     &   47029 & 14.23 & 0.81 &                   4.806 &  0.992 &  1.00 & $-$0.18 \nl   
 85503 & IRC+30218, BS~3905,  $\mu$~Leo & 1.21 &   48455 & 24.52 & 0.87 &    3.878 &  1.222 &  1.13 &  0.17 \nl     
 85859 & IRC$-$30157, BS~3919    & 2.11   &   48559 &  9.71 & 0.66 &         4.866 &  1.199 &  1.19 & $-$0.03 \nl   
 88284 & IRC$-$10233, BS~3994  & 1.41   &   49841 & 28.44 & 0.96 &           3.610 &  1.007 &  0.96 &  0.05 \nl     
 90537 & IRC+40219, BS~4100  & 2.16   &   51233 & 22.34 & 0.87 &             4.196 &  0.908 &  0.89 &  0.00 \nl     
 91612 & IRC+10232  & 2.90  &   51775 & 10.23 & 0.78 &                       5.069 &  0.921 &  0.95 & $-$0.25 \nl   
 92214 & IRC$-$20216  & 2.66   &   52085 & 14.52 & 0.89 &                    4.910 &  0.922 &  0.85 & $-$0.22 \nl   
 92424 & IRC+70097    & 2.51  &   52353 & 14.58 & 0.54 &                     5.119 &  1.207 &  1.17 &  0.06 \nl     
 93813 & IRC$-$20217, BS~4232  & 0.27   &   52943 & 23.54 & 0.81 &           3.109 &  1.232 &  1.22 & $-$0.30 \nl   
 94264 & IRC+30226, BS~4247  & 1.38    &   53229 & 33.40 & 0.78 &            3.789 &  1.040 &  1.07 & $-$0.20 \nl   
 94600 & IRC+30228    & 2.45    &   53426 & 13.33 & 0.75 &                   5.024 &  1.101 &  1.06 & $-$0.26 \nl   
 95272 & IRC$-$20221, BS~4287  & 1.70    &   53740 & 18.71 & 1.03 &          4.080 &  1.079 &  1.06 & $-$0.22 \nl   
 95345 & IRC~~00199  & 2.24    &   53807 &  9.54 & 0.94 &                    4.838 &  1.144 &  1.13 & $-$0.28 \nl   
 96833 & IRC+40224, BS~4335  & 0.42   &   54539 & 22.21 & 0.68 &             3.003 &  1.144 &  1.09 & $-$0.13 \nl   
100407 & IRC$-$30178, BS~4450  & 1.46   &   56343 & 25.23 & 0.83 &           3.540 &  0.947 &  0.92 & $-$0.04 \nl   
100920 & IRC~~00209, BS~4471   & 2.01    &   56647 & 18.31 & 0.89 &          4.304 &  0.983 &  0.98 & $-$0.34 \nl   
102070 & IRC$-$20231,     & 2.54   &   57283 &  9.31 & 0.81 &                4.715 &  0.958 &  0.94 & $-$0.11 \nl   
102224 & IRC+50213, BS~4518  & 0.94   &   57399 & 16.64 & 0.60 &             3.686 &  1.181 &  1.15 & $-$0.44 \nl   
104979 & IRC+10250, BS~4608  & 1.88   &   58948 & 19.08 & 0.77 &             4.123 &  0.967 &  0.96 & $-$0.51 \nl   
106760 & IRC+30236, & 2.33    &   59856 & 10.65 & 0.93 &                     4.987 &  1.140 &  1.12 & $-$0.29 \nl   
107328 & IRC~~00215, BS~4695  & 2.19   &   60172 & 11.43 & 0.82 &            4.970 &  1.172 &  1.19 & $-$0.48 \nl   
108225 & IRC+40234  & 2.89    &   60646 & 14.35 & 0.60 &                     5.014 &  0.955 &  0.94 & $-$0.11 \nl   
109379 & IRC$-$20240, BS~4786, $\beta$~Crv & 0.69 & 61359 & 23.34 & 0.80 &   2.651 &  0.893 &  0.88 & $-$0.11 \nl   
113226 & IRC+10261, BS~4932, $\epsilon$~Vir & 0.78 & 63608 & 31.90 & 0.87 &  2.849 &  0.934 &  0.83 &  0.15 \nl     
114038 & IRC$-$10276  & 2.59   &   64078 & 10.66 & 0.84 &                    5.149 &  1.138 &  1.09 & $-$0.04 \nl   
114149 & IRC$-$20247  & 2.55     &   64166 & 14.14 & 0.80 &                  4.942 &  1.048 &  1.02 & $-$0.19 \nl   
115310 & IRC$-$30203  & 2.91    &   64803 & 13.21 & 0.76 &                   5.095 &  0.959 &  0.95 & $-$0.15 \nl   
115659 & IRC$-$20249, BS~5020  & 0.97  &  64962 & 24.69 & 0.70 &             2.993 &  0.920 &  0.90 & $-$0.12 \nl   
116713 &     &   &   65535 & 15.73 & 0.76 &                                  5.114 &  1.181 &  1.14 & $-$0.25 \nl   
116976 & IRC$-$20253  & 2.41    &   65639 & 12.63 & 0.75 &                   4.757 &  1.096 &  1.02 &  0.02 \nl     
117818 & IRC$-$10289  & 2.96   &   66098 & 12.36 & 0.78 &                    5.210 &  0.964 &  0.95 & $-$0.40 \nl   
119425 & IRC~~00236  & 2.85    &   66936 & 15.01 & 0.86 &                    5.352 &  1.091 &  1.07 & $-$0.01 \nl   
120452 & IRC$-$20261  & 2.58    &   67494 & 13.48 & 0.72 &                   4.959 &  1.059 &  1.09 & $-$0.03 \nl   
123123 & IRC$-$30213, BS~5287  & 0.72   &   68895 & 32.17 & 0.77 &           3.247 &  1.091 &  1.10 & $-$0.16 \nl   
124206 & IRC$-$30216  & 2.39    &   69415 & 14.22 & 0.98 &                   5.075 &  1.130 &  1.16 & $-$0.14 \nl   
124679 & IRC+10278  & 2.97    &   69612 & 12.79 & 0.87 &                     5.294 &  1.007 &  1.04 & $-$0.26 \nl   
125351 & IRC+40255  & 2.44     &   69879 & 14.63 & 0.69 &                    4.796 &  1.057 &  1.00 & $-$0.13 \nl   
125454 & IRC~~00240  & 2.73     &   70012 & 11.90 & 0.95 &                   5.138 &  1.023 &  1.00 & $-$0.22 \nl   
125560 & IRC+20272  & 2.13     &   70027 & 17.12 & 0.68 &                    4.838 &  1.228 &  1.16 &  0.00 \nl     
127665 & IRC+30259, BS~5429, $\rho$~Boo & 0.62 & 71053 & 21.92 & 0.81 &      3.574 &  1.298 &  1.22 & $-$0.17 \nl   
129972 & IRC+20274, BS~5502    & 2.42    &   72125 & 14.48 & 0.79 &          4.604 &  0.972 &  0.94 & $-$0.10 \nl   
130259 &   &       &   72357 &  9.84 & 0.76 &                                5.225 &  0.938 &  0.93 & $-$0.24 \nl   
131111 & IRC+40262   & 2.93  &   72582 & 16.31 & 0.61 &                      5.472 &  1.030 &  1.10 & $-$0.29 \nl   
130952 & IRC~~00251  & 2.60        &   72631 & 15.09 & 1.09 &                4.934 &  0.988 &  0.97 & $-$0.39 \nl   
132132 & IRC~~00254  & 2.92     &   73193 & 11.06 & 0.87 &                   5.509 &  1.131 &  1.09 & $-$0.07 \nl   
133165 & IRC~~00259, BS~5601  & 1.97    &   73620 & 17.78 & 0.90 &           4.390 &  1.026 &  1.04 & $-$0.22 \nl   
133582 & IRC+30268, BS~5616  & 1.63    &   73745 & 13.04 & 0.68 &            4.516 &  1.240 &  1.23 & $-$0.35 \nl   
134190 & IRC+50240, BS~5635   & 2.89   &   73909 & 12.53 & 0.53 &            5.240 &  0.958 &  0.95 & $-$0.49 \nl   
135722 & IRC+30271, BS~5681, $\delta$~Boo & 1.18 & 74666 & 27.94 & 0.61 &    3.461 &  0.961 &  0.96 & $-$0.44 \nl   
136514 & IRC~~00264   & 2.64   &   75119 & 13.38 & 0.86 &                    5.352 &  1.191 &  1.18 & $-$0.14 \nl   
136479 &  &        &   75127 & 12.14 & 0.89 &                                5.541 &  1.047 &  0.99 &  0.06 \nl     
137759 & IRC+60233, BS~5744   & 0.70   &   75458 & 31.92 & 0.51 &            3.290 &  1.166 &  1.07 &  0.03 \nl     
138562 &   &   &   76133 & 12.06 & 0.72 &                                    5.497 &  1.092 &  1.05 & $-$0.13 \nl   
138905 & IRC$-$10323, BS~5787  & 1.51  &   76333 & 21.42 & 0.90 &            3.914 &  1.007 &  1.02 & $-$0.42 \nl   
139195 & 16 Ser  &     &   76425 & 13.89 & 0.70 &                            5.264 &  0.925 &  0.92 & $-$0.17 \nl   
139641 &   &    &   76534 & 20.00 & 0.56 &                                   5.245 &  0.886 &  0.89 & $-$0.55 \nl   
139521 &  &   &   76705 & 13.32 & 0.81 &                                     4.663 &  0.964 &  0.97 & $-$0.34 \nl   
140573 & IRC+10294, BS~5854, $\alpha$~Ser & 0.07 & 77070 & 44.54 & 0.71 &    2.634 &  1.167 &  1.09 & 0.03 \nl      
141714 & IRC+30279, $\delta$~CrB  & 2.71    &   77512 & 19.71 & 0.73 &       4.594 &  0.794 &  0.82 & $-$0.32 \nl   
141680 & IRC~~00273   & 2.81     &   77578 & 12.40 & 0.73 &                  5.207 &  1.019 &  1.01 & $-$0.28 \nl   
142198 & IRC$-$20296, BS~5908  & 1.78   &   77853 & 20.02 & 0.88 &           4.127 &  1.003 &  1.02 & $-$0.31 \nl   
142980 & IRC+10299   & 2.78  &   78132 & 14.36 & 0.80 &                      5.536 &  1.141 &  1.10 &  0.06 \nl     
143107 & IRC+30280, BS~5947, $\epsilon$~CrB & 1.29 & 78159 & 14.20 & 0.70 &  4.142 &  1.231 &  1.17 & $-$0.32 \nl   
143666 & IRC+20290  & 2.78  &   78481 & 10.58 & 0.85 &                       5.102 &  0.992 &  0.97 & $-$0.31 \nl   
143787 & IRC$-$30252  & 2.18  &   78650 & 15.39 & 0.91 &                     4.956 &  1.234 &  1.25 & $-$0.14 \nl   
145328 & IRC+40278, BS~6018   & 2.40   &   79119 & 28.84 & 0.54 &            4.730 &  1.015 &  1.00 & $-$0.20 \nl   
145250 & IRC$-$30256 & 2.43   &   79302 & 12.77 & 0.92 &                     5.091 &  1.131 &  1.09 & $-$0.36 \nl   
146791 & IRC~~00282, BS~6075  & 0.98   &   79882 & 30.34 & 0.79 &            3.230 &  0.966 &  0.96 & $-$0.25 \nl   
147677 & IRC+30287  & 2.59   &   80181 & 17.76 & 0.63 &                      4.857 &  0.970 &  0.93 & $-$0.08 \nl   
148387 & IRC+60242, BS~6132  & 0.61   &   80331 & 37.18 & 0.45 &             2.732 &  0.910 &  0.84 & $-$0.21 \nl   
148786 & IRC$-$20318, BS~6147, $\phi$~Oph & 2.27 & 80894 & 15.53 & 0.77 &    4.286 &  0.924 &  0.89 &  0.08 \nl     
150449 & IRC+60245  & 2.72   &   81437 & 12.60 & 0.46 &                      5.282 &  1.055 &  1.02 & $-$0.12 \nl   
150997 & IRC+40287,  BS~6220  & 1.33   &   81833 & 29.11 & 0.52 &            3.480 &  0.916 &  0.89 & $-$0.37 \nl   
151680 &    BS~6241      & $-$0.25   &   82396 & 49.85 & 0.81 &              2.288 &  1.144 &  1.10 & $-$0.17 \nl   
152815 &  &        &   82764 & 12.80 & 0.77 &                                5.386 &  0.966 &  0.95 & $-$0.26 \nl   
153210 & IRC+10315, BS~6299, 27~Oph & 0.66  & 83000 & 37.99 & 0.75 &         3.188 &  1.160 &  1.10 & $-$0.03 \nl   
156266 & IRC~~00299  & 2.16     &   84514 & 14.97 & 0.72 &                   4.716 &  1.119 &  1.09 & $-$0.03 \nl   
159966 & IRC+70140  & 2.62     &   85805 & 15.02 & 0.54 &                    5.073 &  1.077 &  1.04 & $-$0.20 \nl   
161096 & IRC~~00317, BS~6603, $\beta$~Oph & 0.23 & 86742 & 39.78 & 0.75 &    2.758 &  1.168 &  1.10 &  0.02 \nl     
162211 & IRC+30316  & 2.46     &   87194 & 15.94 & 0.61 &                    5.093 &  1.141 &  1.09 & $-$0.03 \nl   
163217 & IRC+40305  & 2.35     &   87563 &  8.97 & 0.53 &                    5.174 &  1.166 &  1.13 & $-$0.12 \nl   
163588 & IRC+60253, BS~6688  & 1.03   &   87585 & 29.26 & 0.49 &             3.731 &  1.177 &  1.11 & $-$0.09 \nl   
163532 & IRC~~00331    & 2.72     &   87847 &  7.66 & 0.71 &                 5.441 &  1.162 &  1.12 & $-$0.16 \nl   
163993 & IRC+30324, BS~6703  & 1.59     &   87933 & 24.12 & 0.52 &           3.702 &  0.935 &  0.89 & $-$0.10 \nl   
163917 & IRC$-$10387, BS~6698  & 1.14     &   88048 & 21.35 & 0.79 &         3.317 &  0.987 &  0.95 & 0.02 \nl      
165135 & IRC$-$30353, BS~6746, $\gamma$~Sgr & 0.67 & 88635 & 33.94 & 0.87 &  2.984 &  0.981 & 0.99 & $-$0.36 \nl    
165760 & IRC+10350  & 2.43 &   88765 & 13.71 & 0.82 &                        4.645 &  0.951 &  0.92 & $-$0.10 \nl   
166208 & IRC+40309     & 2.99     &   88788 &  8.98 & 0.50 &                 4.996 &  0.913 &  0.91 & $-$0.06 \nl   
166640 &   &    &   89008 &  7.52 & 0.57 &                                   5.571 &  0.915 &  0.91 & $-$0.14 \nl   
166460 & IRC~~00339  & 2.75   &   89065 &  7.91 & 0.74 &                     5.501 &  1.204 &  1.16 & $-$0.03 \nl   
166464 & IRC$-$20443  & 2.55   &   89153 & 13.35 & 0.81 &                    4.957 &  1.055 &  1.02 & $-$0.18 \nl   
168387 & IRC+10355 & 2.82   &   89772 & 22.24 & 0.82 &                       5.406 &  1.084 &  1.05 &  0.05 \nl     
168775 & IRC+40313, BS~6872     & 1.77  &   89826 & 13.71 & 0.56 &           4.335 &  1.162 &  1.10 & $-$0.09 \nl   
168656 & IRC~~00346  & 2.75   &   89918 & 12.11 & 0.83 &                     4.846 &  0.911 &  0.90 & $-$0.21 \nl   
168723 & IRC~~00347, BS~6869, $\eta$~Ser & 1.04 & 89962 & 52.81 & 0.75 &     3.234 &  0.941 &  0.96 & $-$0.42 \nl   
169156 & IRC$-$10417   & 2.43   &   90135 & 17.08 & 0.87 &                   4.663 &  0.932 &  0.94 & $-$0.17 \nl   
169414 & IRC+20364, BS~6895  & 1.12    &   90139 & 25.40 & 0.65 &            3.854 &  1.168 &  1.13 & $-$0.16 \nl   
170693 & IRC+70145   & 1.95     &   90344 & 10.28 & 0.48 &                   4.820 &  1.179 &  1.16 & $-$0.44 \nl   
169916 & IRC$-$30386, BS~6913, 22~Sgr & 0.40 & 90496 & 42.20 & 0.90 &        2.817 &  1.025 &  1.04 & $-$0.20 \nl   
170474 &   &      &   90642 & 13.81 & 0.78 &                                 5.381 &  0.961 &  0.95 & $-$0.08 \nl   
171391 &   &     &   91105 & 11.25 & 0.78 &                                  5.116 &  0.926 &  0.92 & $-$0.07 \nl   
173780 & IRC+30342, BS~7064  & 2.01   &   92088 & 12.96 & 0.54 &             4.833 &  1.199 &  1.16 & $-$0.12 \nl   
175535 & IRC+50286   & 2.73   &   92689 &  9.67 & 0.57 &                     4.917 &  0.903 &  0.88 & $-$0.14 \nl   
176524 & IRC+70147  & 2.24   &   92782 &  9.47 & 0.46 &                      4.822 &  1.151 &  1.10 & $-$0.12 \nl   
175515 & IRC+10386  & 3.12   &   92872 & 11.14 & 0.79 &                      5.584 &  1.041 &  1.01 & $-$0.26 \nl   
175751 & IRC$-$10480  & 2.33    &   93026 & 15.77 & 0.89 &                   4.829 &  1.057 &  1.03 & $-$0.11 \nl   
176411 & IRC+10392, BS~7176, $\epsilon$~Aql & 2.12 & 93244 & 21.22 & 0.77 &  4.024 &  1.082 &  1.00 &   0.00 \nl    
176678 & IRC$-$10483, BS~7193  & 1.51   &  93429 & 21.95 & 0.92 &            4.017 &  1.079 &  1.08 & $-$0.19 \nl   
177241 & IRC$-$20536, BS~7217  & 1.48   &   93683 & 23.49 & 0.78 &           3.755 &  1.012 &  0.98 & $-$0.11 \nl   
177716 & IRC$-$30401, BS~7234  & 0.64   &   93864 & 27.09 & 1.48 &           3.324 &  1.169 &  1.15 & $-$0.17 \nl   
180711 & IRC+70150, BS~7310  & 0.78   &   94376 & 32.54 & 0.46 &             3.072 &  0.990 &  0.94 & $-$0.27 \nl   
181984 & IRC+70153, BS~7352  & 1.74   &   94648 & 21.73 & 0.47 &             4.445 &  1.257 &  1.15 &   0.12 \nl    
181276 & IRC+50291, BS~7328  & 1.64   &   94779 & 26.48 & 0.49 &             3.795 &  0.950 &  0.85 & $-$0.08 \nl   
182762 & IRC+20402, & 2.69   &   95498 & 13.78 & 0.74 &                      5.136 &  0.999 &  0.98 & $-$0.20 \nl   
184406 & IRC+10430,  BS~7429  & 1.79   &   96229 & 29.50 & 0.78 &            4.449 &  1.176 &  1.14 & $-$0.13 \nl   
185734 & IRC+30380   & 2.45   &   96683 & 13.00 & 0.59 &                     4.680 &  0.971 &  0.89 & $-$0.11 \nl   
186675 & IRC+40361   & 2.73   &   97118 & 11.70 & 0.50 &                     4.891 &  0.948 &  0.94 & $-$0.14 \nl   
186648 & IRC$-$20572 & 2.44   &   97290 & 15.92 & 0.77 &                     4.870 &  1.061 &  1.03 & $-$0.18 \nl   
188119 & IRC+70160, BS~7582  & 1.68   &   97433 & 22.40 & 0.45 &             3.841 &  0.888 &  0.88 & $-$0.47 \nl   
188310 & IRC+10442   & 2.37   &   97938 & 15.96 & 1.01 &                     4.715 &  1.023 &  1.03 & $-$0.32 \nl   
188947 & IRC+30401, BS~7615, $\eta$~Cyg & 1.63  &   98110 & 23.40 & 0.54 &   3.886 &  1.019 &  0.98 & $-$0.09 \nl   
189005 & IRC$-$30422 & 2.73   &   98353 &  9.58 & 0.83 &                     4.837 &  0.882 &  0.91 & $-$0.38 \nl   
190056 & IRC$-$30424 & 2.06   &   98842 & 10.05 & 0.84 &                     4.991 &  1.208 &  1.17 & $-$0.63 \nl   
190608 & IRC+20453  & 2.71   &   98920 & 20.17 & 0.71 &                      5.094 &  1.058 &  1.10 & $-$0.03 \nl   
191277 & IRC+60279  & 2.70   &   98962 & 18.78 & 0.44 &                      5.399 &  1.191 &  1.15 &  0.12 \nl     
192947 & IRC$-$10535, BS~7754    & 1.47  &  100064 & 30.01 & 0.91 &          3.576 &  0.883 &  0.92 & $-$0.18 \nl   
194152 &  &  &  100437 &  7.86 & 0.53 &                                      5.577 &  1.077 &  1.04 &  0.00 \nl     
195135 & IRC~~00476  & 2.32    &  101101 & 17.08 & 0.94 &                    4.913 &  1.160 &  1.12 &  0.03 \nl     
196758 & IRC~~00486 & 2.78   &  101936 & 13.33 & 0.88 &                      5.154 &  1.060 &  1.02 & $-$0.12 \nl   
196737 &     &     &  102014 & 13.64 & 0.79 &                                5.474 &  1.118 &  1.08 & $-$0.02 \nl   
197912 & IRC+30450, BS~7942 & 1.83     &  102453 & 15.84 & 0.62 &            4.219 &  1.051 &  1.01 & $-$0.24 \nl   
197989 & IRC+30451, BS~7949, $\epsilon$~Cyg & 0.11 & 102488 & 45.26 & 0.53 & 2.479 &  1.021 &  1.00 & $-$0.27 \nl   
197964 & IRC+20482, $\gamma$~Del  & 1.82  &  102532 & 32.14 & 1.19 &         4.275 &  1.042 &  1.03 &  0.13 \nl     
198809 & IRC+30458   & 2.97  &  103004 & 15.06 & 0.60 &                      4.559 &  0.835 &  0.87 & $-$0.23 \nl   
199870 &     &  &  103519 & 12.32 & 0.63 &                                   5.553 &  0.973 &  0.96 &  0.00 \nl     
199951 & IRC$-$30439,  & 2.81    &  103738 & 14.59 & 0.79 &                  4.670 &  0.890 &  0.90 & $-$0.23 \nl   
200763 & IRC$-$30440, & 2.79   &  104174 &  9.22 & 0.74 &                    5.204 &  1.104 &  1.07 & $-$0.01 \nl   
201381 & IRC$-$10557, BS~8093  & 2.40    &  104459 & 19.93 & 0.77 &          4.498 &  0.926 &  0.92 & $-$0.15 \nl   
202109 & IRC+30472, BS~8115, $\zeta$~Cyg & 1.08 & 104732 & 21.62 & 0.63 &    3.214 &  0.990 &  0.97 & $-$0.11 \nl   
203344 &     &  &  105411 & 11.63 & 0.73 &                                   5.578 &  1.057 &  1.09 & $-$0.24 \nl   
203504 & IRC+20505, BS~8173  & 1.61     &  105502 & 21.19 & 0.87 &           4.081 &  1.108 &  1.05 & $-$0.14 \nl   
203387 & IRC$-$20599, BS~8167, $\iota$~Cap & 2.25 & 105515 & 15.13 & 0.80 &  4.279 &  0.888 &  0.89 & $-$0.23 \nl   
204381 & IRC$-$20604, BS~8213    & 2.44  &  106039 & 18.18 & 0.89 &          4.498 &  0.889 &  0.89 & $-$0.28 \nl   
205435 & IRC+50385, BS~8252, $\rho$~Cyg & 1.92  &  106481 & 26.20 & 0.51 &   3.982 &  0.885 &  0.94 & $-$0.31 \nl   
205512 & IRC+40486, BS~8255    & 2.48    &  106551 & 12.76 & 0.53 &          4.869 &  1.085 &  1.06 & $-$0.08 \nl   
206067 & IRC~~00506    & 2.84   &  106944 & 13.55 & 0.88 &                   5.100 &  1.034 &  1.01 & $-$0.17 \nl   
206952 & IRC+70175, 11 Cep    & 2.07   &  107119 & 18.55 & 0.46 &            4.552 &  1.108 &  1.07 &  0.04 \nl     
206356 & IRC$-$20606   & 2.98   &  107128 & 13.18 & 0.80 &                   5.240 &  0.991 &  0.97 & $-$0.01 \nl   
206453 & IRC$-$20607   & 2.64   &  107188 & 11.22 & 0.79 &                   4.720 &  0.868 &  0.91 & $-$0.20 \nl   
209396 &  &    &  108868 & 12.02 & 0.80 &                                    5.550 &  0.959 &  0.95 & $-$0.11 \nl   
210960 &  &   &  109786 & 11.59 & 0.98 &                                     5.330 &  0.812 &  0.84 & $-$0.13 \nl   
211391 & IRC$-$10578, BS~8499, $\theta$~Aqr &2.02 &110003 &17.04 & 0.74 &    4.167 &  0.979 &  0.95 &  0.01 \nl     
212271 &  &   &  110529 & 12.00 & 0.77 &                                     5.526 &  0.979 &  0.96 & $-$0.07 \nl   
212496 & IRC+50428, BS~8538, $\beta$~Lac & 2.08 & 110538 & 19.21 & 0.51 &    4.418 &  1.015 &  1.03 & $-$0.39 \nl   
212943 & IRC~~00517, BS~8551  & 2.31  &  110882 & 20.39 & 0.78 &             4.777 &  1.039 &  1.07 & $-$0.37 \nl   
214376 & IRC~~00521,   & 2.63   &  111710 & 13.92 & 0.75 &                   5.038 &  1.140 &  1.10 &  0.14 \nl     
214995 &  &    &  112067 & 12.22 & 0.79 &                                    5.925 &  1.114 &  1.08 & $-$0.09 \nl   
215721 & IRC$-$20622,   & 2.97    &  112529 & 12.26 & 0.87 &                 5.243 &  0.941 &  0.93 & $-$0.43 \nl   
216228 & IRC+70190, BS~8694  & 1.16   &  112724 & 28.27 & 0.52 &             3.497 &  1.053 &  1.06 & $-$0.12 \nl   
216174 & IRC+60368   & 2.64   &  112731 &  8.78 & 0.58 &                     5.428 &  1.167 &  1.13 & $-$0.53 \nl   
216131 & IRC+20537, BS~8684    & 1.38    &  112748 & 27.95 & 0.77 &          3.512 &  0.933 &  0.89 & $-$0.16 \nl   
216646 &  &    &  113084 &  9.63 & 0.79 &                                    5.818 &  1.136 &  1.10 & $-$0.02 \nl   
216718 &  &   &  113184 & 11.13 & 0.95 &                                     5.723 &  0.881 &  0.89 &  0.04 \nl     
216763 & IRC$-$30457   & 2.05    &  113246 & 19.14 & 0.87 &                  4.200 &  0.952 &  0.96 & $-$0.31 \nl   
217264 &  &     &  113521 & 11.63 & 0.93 &                                   5.428 &  0.982 &  0.96 & $-$0.04 \nl   
218029 & IRC+70192    & 2.48   &  113864 &  8.48 & 0.52 &                    5.252 &  1.248 &  1.21 &  0.07 \nl     
218031 & IRC+50457   & 2.21   &  113919 & 18.20 & 0.57 &                     4.640 &  1.058 &  1.02 & $-$0.20 \nl   
218240 & IRC$-$20628    & 2.40   &  114119 & 17.32 & 0.88 &                  4.483 &  0.892 &  0.92 & $-$0.14 \nl   
218658 & IRC+80056   & 2.46   &  114222 & 14.83 & 0.62 &                     4.406 &  0.802 &  0.84 &  0.01 \nl     
219139 &  &   &  114641 &  9.72 & 0.77 &                                     5.848 &  1.004 &  0.98 & $-$0.33 \nl   
219449 & IRC$-$10596, BS~8841  & 1.78   &  114855 & 21.97 & 0.89 &           4.239 &  1.107 &  1.06 & $-$0.14 \nl   
219615 & IRC~~00528, BS~8852  & 1.44     &  114971 & 24.92 & 0.89 &          3.703 &  0.916 &  0.97 & $-$0.61 \nl   
219916 & IRC+70194     & 2.77   &  115088 & 15.48 & 0.55 &                   4.749 &  0.836 &  0.86 & $-$0.07 \nl   
219784 & IRC$-$30468  & 1.96  &  115102 & 18.24 & 0.80 &                     4.405 &  1.109 &  1.08 & $-$0.22 \nl   
220009 & IRC+10532   & 2.07   &  115227 &  9.56 & 0.95 &                     5.052 &  1.204 &  1.16 & $-$0.63 \nl   
220321 & IRC$-$20633, BS~8892  & 1.37  &  115438 & 20.14 & 0.72 &            3.960 &  1.082 &  1.10 & $-$0.40 \nl   
220954 & IRC+10535, BS~8916     & 1.87   &  115830 & 20.54 & 0.80 &          4.273 &  1.062 &  1.03 & $-$0.12 \nl   
221115 & IRC+10536, BS~8923   & 2.40    &  115919 & 18.34 & 0.74 &           4.537 &  0.939 &  0.93 & $-$0.03 \nl   
221345 & IRC+40539  & 2.73    &  116076 & 13.09 & 0.71 &                     5.215 &  1.029 &  1.00 & $-$0.36 \nl   
222093 &     &   &  116591 & 11.50 & 0.89 &                                  5.664 &  1.025 &  1.00 & $-$0.25 \nl   
222493 &  &    &  116853 &  8.60 & 0.86 &                                    5.893 &  0.984 &  0.97 & $-$0.02 \nl   
222842 & IRC+30516  & 2.69  &  117073 & 13.91 & 0.75 &                       4.927 &  0.935 &  0.93 & $-$0.16 \nl   
223170 &    &     &  117314 &  9.42 & 0.83 &                                 5.740 &  1.068 &  1.03 & $-$0.14 \nl   
223252 &   &    &  117375 & 11.19 & 0.85 &                                   5.494 &  0.941 &  0.96 & $-$0.11 \nl   
224533 & IRC~~00536  & 2.69   &  118209 & 14.58 & 0.83 &                     4.879 &  0.930 &  0.92 & $-$0.13 \nl   
\enddata
\tablenotetext{A}{Prefixes: IRC = Infrared Catalog (aka TMSS), some IRC names appear to be missing
$\pm$ signs, but are listed as found in the CIO; BS = Yale Bright Star Catalog, see also the original listing
in the Harvard Revised Catalog (the prefix BS is used in the CIO).}
\end{deluxetable}
\end{landscape}

\end{document}